\documentclass{article}%
\usepackage{amsmath}
\usepackage{amsfonts}
\usepackage{amssymb}
\usepackage{graphicx}%
\begin{document}

\begin{center}
\bigskip CHARGE FLUCTUATION OF DUST GRAINS AND ITS IMPACT ON DUSTY WAVE.PROPAGATION

Barbara Atamaniuk and Krzysztof \.{Z}uchowski

I{\small nstitute of Fundamental Technological Research, Polish Academy of
Sciences}

\bigskip
\end{center}

{\small In this paper we consider the influence of dust charge fluctuations on
damping of the dust-ion-acoustic waves. Fluid approximation of longitudinal
electrostatic waves in unmagnetized plasmas is considered. We show that for a
weak acoustic wave the attenuation depends on a phenomenological charging
coefficient.}

\bigskip

\ \ \ \textbf{\ Introduction}

\noindent\ \ \ Plasmas with dust grains are of interest both for the cosmic
space as well as for the laboratory plasmas. Examples include cometary
environments, planetary rings, the interstellar medium, the Earth's and other
planets magnetospheres \cite{Goertz89}. Dust has been found to be a
determinant component of rarefied plasmas used in the microelectronic
processing industry \cite{Merlino}, and it may also be present in the limiter
regions of fusion plasmas due to sputtering of the carbon by energetic
particles. It is interesting to note the recent flurry of activity in the
dusty plasma research. It has been driven largely by discoveries of the role
of dust in quite different settings: the ring of Saturn \cite{Goertz89} and
the plasma processing device \cite{Merlino}. Dusty plasmas contain, beside
positive ions and electrons, large particles usually negatively charged. They
are conglomerations of the ions, electrons and neutral particles. These large
particles, to be called grains, have atomic numbers $Z_{d}$ in the range of
$10^{4}-10^{6}$ and their mass $m_{d}$ can be equal to $10^{6}$ of the
\ proton mass or even much more. In the considered dusty plasmas, the size of
grains is small compared with average distance between the grains.The ratio of
charge to mass for a given component of plasma determines its dynamics. We
note that for dusty plasmas, ratio of electrical charges of grains to their
masses is usually much smaller than in the case of multispecies plasmas with
negative ions and hence, here comes the first of the crucial differences
between multispecies plasmas with negative ions and dusty plasmas. Because
dynamics of the dusty plasma components, electrons, positive ions and dust
grains is quite different in the time and length scales considered here, then
the equations for these components of the dusty plasma may be different. If it
is assumed that all grains have equal masses and charges steady in time,
therefore the dust-ion-acoustic and dust-acoustic dispersion relations are
obtained on the basis of fluid \cite{Shukla92}, \cite{Shukla92a} or kinetic
\cite{Turski99} models. In this case we have assumed, for simplicity, that all
grains have equal masses and charges, but charges are not constant in time -
they may fluctuate in time. The dust charges are not really independent of the
variations in the plasma potentials. Here, even in the fluid theory,
appear\ the crucial differences between the ordinary multispecies plasmas and
the dusty plasmas. All modes will influence the charging mechanism, and
feedback will lead to several new interesting and unexpected phenomena. The
charging of the grains depends on local plasma characteristics. If the waves
disturb these characteristic, then charging of the grains is affected and the
grain charge is modified, with a resulting feedback on the wave mode.

The simplest cases to deal with are the parallel electrostatic modes in an
unmagnetized plasma. Then the electric field $E$ is one-dimensional and may be
represented by the electric potential $\phi$: $E=-\partial\phi/\partial x$. We
consider a problem when the temperature of electrons $T_{e}$ is much greater
than the temperature of ions $T_{i}$: $T_{e}\gg T_{i}$. In such simplified
situations, fluctuations in time of the number density of electrons $\delta
n_{e}$ can occur due to the grains of the dust loosing or picking up some
electrons. Dust charge fluctuations in time give rise to purely damped
acoustic modes when the streams of particles are absent. Here we solve the
continuity equations in approximation for the source term vanishing at equilibrium.

\section{Fluctuation of dust grains in dusty plasmas}

\bigskip\noindent

As a result of fluctuating dust charges in dusty plasmas, many new
problems\ can appear which are in partly treatment by Verheest
\cite{Verheest00}. We consider a specific problem when the temperature of
electrons is much greater than the temperature of ions and we also assume that
the mass of grains with fluctuating charges may be approximated by constant
values. In this case the continuity equations for specimens of dusty plasmas
can be written in the form:%
\[
\partial n_{d}/\partial t+\partial(n_{d}u_{d})/\partial x=0,
\]%
\begin{equation}
\partial n_{i}/\partial t+\partial(n_{i}u_{i})/\partial x=0, \tag{2.1}%
\label{2.1}%
\end{equation}%
\[
\partial n_{e/}\partial t+\partial(n_{e}u_{e})/\partial x=S_{e}.
\]

\bigskip Due to the possible fluctuations of the dust charges we can express
the conservation of charge in the dusty plasma by:%

\begin{equation}
\frac{\partial}{\partial t}\left(  -n_{e}e+n_{d}q_{d}+n_{i}e\right)
+\frac{\partial}{\partial x}\left(  -n_{e}eu_{e}+n_{d}q_{d}u_{d}+n_{i}%
eu_{i}\right)  =0, \tag{2.2}\label{2.2}%
\end{equation}
where $q_{d}$ is the charge of grain of dust. This can be rewritten with the
help of the continuity equation (2.1 - 2.3) as:%

\begin{equation}
n_{d}\left(  \frac{\partial}{\partial t}+u_{d}\frac{\partial}{\partial
x}\right)  q_{d}=e\,S_{e}. \tag{2.3}\label{2.3}%
\end{equation}
On the other hand, the charge of grain of dust fluctuation is given by:
\begin{equation}
\frac{dq_{d}}{dt}=\left(  \frac{\partial}{\partial t}+u_{d}\frac{\partial
}{\partial x}\right)  q_{d}=I_{i}\left(  n_{i,}q_{d}\right)  +I_{e}\left(
n_{e},q_{d}\right)  , \tag{2.4}\label{2.4}%
\end{equation}
where $I_{i}\left(  n_{i},q_{d}\right)  $ and $I_{e}\left(  n_{e}%
,q_{d}\right)  $ are the ionic and electronic charging current, respectively.
When we combine (2. 3) and (2.4), we get
\begin{equation}
eS_{e}=n_{d}I_{e}\left(  n_{e},q_{d}\right)  +n_{d}I_{i}\left(  n_{i,}%
q_{d}\right)  . \tag{2.5}\label{2.5}%
\end{equation}
In equilibrium dusty plasma, the total charging current vanishes:
\begin{equation}
I_{i0}+I_{e0}=0, \tag{2.6}\label{2.6}%
\end{equation}
where $I_{i0}$ and $I_{e0}$ denotes the equilibrium charging current for ions
and electrons, respectively. Therefore we can expand (2.5) as a function of
$n_{e}$, $q_{d}$ and $n_{d}$ using (2.6) and \ hence in linear approximation
for $S_{e}$ vanishing at equilibrium, it is given by:
\begin{equation}
S_{e}=-\nu_{e}\delta n_{e}-\mu_{e}\delta q_{d}, \tag{2.7}\label{2.7}%
\end{equation}
where $\nu_{e}$, $\mu_{e}$ denotes charging fluctuation coefficients while
$\delta n_{e}$ \ and $\delta q_{d}$ denotes fluctuation electron number
density and \ fluctuation charges of grains from their equilibrium values respectively.

\section{ Dumping dust-acoustic wave}

\bigskip

Now we add to the continuity equations (2.1), (\ref{2.3}) \ the equations of
motion written explicitly for\ a three-component dusty plasma.%
\[
\left(  \frac{\partial}{\partial t}+u_{d}\frac{\partial}{\partial x}\right)
u_{d}=-\frac{q_{d}}{m_{d}}\frac{\partial\phi}{\partial x},
\]%
\[
\left(  \frac{\partial}{\partial t}+u_{i}\frac{\partial}{\partial x}\right)
u_{i}+\frac{c_{si}^{2}}{n_{i}}\frac{\partial n_{i}}{\partial x}=-\frac
{e}{m_{i}}\frac{\partial\phi}{\partial x},
\]

\[
\left(  \frac{\partial}{\partial t}+u_{e}\frac{\partial}{\partial x}\right)
u_{e}+\frac{c_{se}^{2}}{n_{e}}\frac{\partial n_{e}}{\partial x}=\frac{e}%
{m_{i}}\frac{\partial\phi}{\partial x}.
\]

\noindent Here $\phi$, $\,c_{se}$, $c_{si}\left(  c_{si}^{2}=\frac{k_{B}T_{i}%
}{m_{i}}\text{, }c_{se}^{2}=\frac{k_{B}T_{e}}{m_{e}}\text{-if the electron
obeys an ideal gas law}\right)  $, $m_{i}$, $m_{e}$, $k_{B}$, $T_{e}$, $T_{i}%
$-are electric potential, thermal velocity of electrons and ion, mass of ion
and electron, Boltzmann constant, temperature of electrons and ions,
respectively. The fluid equations are supplemented by Poisson's equation:
\begin{equation}
\varepsilon_{0}\frac{\partial^{2}\phi}{\partial x^{2}}=en_{e}-en_{i}%
-q_{d}n_{d}, \tag{3.2}\label{3.2}%
\end{equation}
where $\varepsilon_{0}$ denotes free space permittivity.

Assuming sufficiently small disturbances of plasma equilibrium, we linearize
the relevant equations around the equilibrium and Fourier transform. Taking
into account the conservation\ of total charge and the fact that the phase
velocities of wave are smaller than the thermal velocity of electrons and
larger than the thermal velocities of ions and dust:$\ c_{sd},c_{si}\ll
\frac{\omega}{k}\ll c_{se},$where $\omega$ and $k$ denotes the wave frequency
and wave number, we obtain -\ after some calculations -\ dispersion relation
for the dust-ion-acoustic waves including the charge fluctuation of dust:
\begin{equation}
\omega=\omega_{0}-i\frac{k^{2}\lambda_{De}^{2}\omega_{pi}^{2}}{2\omega_{0}%
^{2}\left(  1+k^{2}\lambda_{De}^{2}\right)  ^{2}}\nu_{e}, \tag{3.3}\label{3.3}%
\end{equation}
where
\begin{equation}
\omega_{0}=\sqrt{\frac{k^{2}\lambda_{De}^{2}\omega_{pi}^{2}}{1+k^{2}%
\lambda_{De}^{2}}+k^{2}c_{si}^{2}}, \tag{3.4}\label{3.4}%
\end{equation}

\bigskip$\lambda_{De}=\sqrt{\varepsilon_{0}k_{B}T_{e}/N_{0e}e^{2}}$ is the
electron Debye length, $\omega_{pi}=\sqrt{N_{0i}e^{2}/\varepsilon_{0}m_{i}}$
is the ion plasma frequency. The equilibrium number density is related by
quasi-neutrality of dusty plasma relation: \ \ \ $N_{0i}=N_{0e}+Z_{d}N_{0d},$
where $q_{d}=eZ_{d}$ and \ $Z_{d}$ - is the atomic number grain of dust.

For $k\rightarrow0$ we have the following dispersion relation for the
dust-ion-acoustic waves (DIAW):%

\begin{equation}
\omega=\omega_{0}-i\nu_{e}\frac{\lambda_{De}^{2}}{2\left(  \lambda_{De}%
^{2}+\lambda_{Di}^{2}\right)  }\approx kc_{s}-\frac{1}{2}i\nu_{e}.
\tag{3.5}\label{3.5}%
\end{equation}
This equation describes the dumped dust-ion-acoustic waves \ including the
charge fluctuation. In our approximation: \ $T_{e}>>T_{i}$, the dumping of
dust-ion- acoustic waves is independent of the parameter $\mu_{e}$.

\section{Conclusions}

\bigskip

The paper deals with a small dust charge fluctuations. In the case considered
here , when the temperature of electrons is much greater than the temperature
of the ions: $T_{e}>>T_{i}$ and $T_{e}$ is not great enough for further
ionization of the ions, we show that attenuation of the acoustic wave depends
only on one phenomenological coefficient $\nu_{e}$. The value of this
coefficient depends mainly on the temperature of electrons.

*\textbf{This research is supported by KBN grant 2PO3B-126- 24,}


\begin{thebibliography}{9}                                                                                                %


\bibitem {Goertz89}C.K. Geortz, Reviews of Geophysics, \textit{Dusty plasmas
in the solar system}, 27,2/May (271-292) 1989.

\bibitem {Merlino}R.L. Merlino, A. Barkan, C. Thompson, N.
D'angelo,\textit{\ Laboratory studies of waves and instabilities in dusty
plasmas, Phys}. Plasmas 5, 1607, 1998

\bibitem {Shukla92}P.K. Shukla, V.P. Silin, \textit{Dust ion-acoustic wave},
Phys.Scr.,45, 508, 1992.

\bibitem {Shukla92a}P.K. Shukla, \textit{Low- frequency modes in dusty
plasmas}, Phys.Scr.,45, 504, 1992.

\bibitem {Turski99}A.J. Turski, B. Atamaniuk and K. \.{Z}uchowski,
\textit{Dusty plasma solitons in Vlasow plasmas}, Arch. Mech.,51,167, 1999.

\bibitem {Verheest00}F. Verheest , \textit{Waves in Dusty Space Plasmas},
Kluwer, Dordrecht 2000.
\end{thebibliography}
\end{document}